\documentclass[twocolumn]{revtex4}

\usepackage{amssymb}
\usepackage{amsmath}
\usepackage{graphicx}
\newcommand{\pa}{\partial}

\newcommand{\be}{\begin{equation}}
\newcommand{\ee}{\end{equation}}
\newcommand{\bea}{\begin{eqnarray}}
\newcommand{\eea}{\end{eqnarray}}

\begin{document}


\title{Double Lynden-Bell Structure of Low-Energy Quasi-Stationary Distributions in the Hamiltonian Mean-Field Model}

\author{Eiji Konishi\footnote{Corresponding author: konishi.eiji.27c@st.kyoto-u.ac.jp} and Masa-aki Sakagami\footnote{sakagami.masaaki.6x@kyoto-u.ac.jp}}
\affiliation{Graduate School of Human and Environment Studies, Kyoto University, Kyoto 606-8501, Japan}


\date{\today}

\begin{abstract}
In the Hamiltonian mean-field model, we study the core-halo structure of low-energy quasi-stationary states under unsteady water-bag type initial conditions.
The core-halo structure results in the superposition of two independent Lynden-Bell distributions.
We examine the completeness of the Lynden-Bell relaxation and the relaxation between these two Lynden-Bell distributions.
\end{abstract}

\pacs{}
\keywords{}

\maketitle

\section{Introduction}
  One of the most interesting features of long-range systems is the existence of a non-equilibrium and long-lived quasi-stationary state (QSS).\cite{Book,ReviewI,BMR,MRS}
The life time of this QSS diverges with the number of particles, and the QSS temporally separates the Vlasov collisionless regime\cite{BH} from the collisional regime of the system.\cite{Book,ReviewI}
There have been many investigations revealing the various type structures of QSSs appearing in self-gravitating systems, non-neutral plasma systems, and models in diverse dimensions.\cite{Book,ReviewI,ReviewII}
Recently, it has been a great challenge to understand the non-equilibrium core-halo structure in QSSs beyond the paradigmatic Lynden-Bell model of QSSs in such long-range systems.\cite{Book,ReviewII,LPT,TLPR,TLP,LB}

The Hamiltonian mean-field (HMF) model is a widely studied benchmark model for long-range systems.\cite{HMF,ReviewI,Book,ReviewII}
In the HMF model, Pakter and Levin explained the origin of their proposed uniform ansatz for the core-halo structure of {{low-energy QSSs}} by the parametric resonance of the initial (first one or two periods of) strong oscillation of magnetization.\cite{PL}
The resonant particles form the high-energy {\it{halo}} and, at the same time, the remaining particles condense into the low-energy dense {\it{core}}. The latter Fermi degeneration-like phenomenon is due to the Vlasov incompressibility.
The resultant distribution considerably deviates from the Lynden-Bell equilibrium one\cite{AFBCDR,ACFR,AFRY}.
Benetti {\it{et al}}.\cite{BTPL} advanced Pakter and Levin's argument to ergodicity breaking and introduced the generalized virial condition (GVC) on the HMF model, deviation from which reflects the deviation from the Lynden-Bell equilibrium.

Pakter and Levin modeled the core-halo structure in QSSs in the HMF model as the {\it{attachment}} of two energy water-bag distributions.\cite{PL}
In contrast, we investigate the possibility of describing the core-halo structure in QSSs as a {\it{superposition}} of two independent Lynden-Bell energy distributions, that is, a {\it{double Lynden-Bell structure}}.
Its halo matches the deviation from the Lynden-Bell equilibrium in general cases and, remarkably, this structure conserves ergodicity independently for its core and halo.
In this paper, we intend to corroborate this scenario in the HMF model at low energies per particle and examine the completeness of the collisionless relaxation\cite{Chavanis1,CSR,Chavanis2} in two ways: by using the Lynden-Bell entropy and by using the double Lynden-Bell entropy.

The rest of this paper is organized as follows.
In Sec.II, we propose the double Lynden-Bell scenario in the HMF model.
In Sec.III, we corroborate this scenario at low energies per particle with $N$-body simulation results.
In Sec.IV, we examine the completeness of the collisionless relaxation by using the double Lynden-Bell entropy.
In Sec.V, we summarize the overall results and indicate some open issues.

\section{Double Lynden-Bell scenario}
The $N$-body HMF model treats $N$ identical particles with unit mass, $XY$ interacting on a circle of radius one.
Their dynamics is governed by the Hamiltonian
\begin{equation}
H=\sum_{j=1}^N \frac{p_j^2}{2}+\frac{1}{2N}\sum_{j,k=1}^N[1-\cos (\theta_j-\theta_k)]\;,
\end{equation}
where angle $\theta_j$ is the orientation of the $j$-th particle and $p_j$ is its conjugate momentum.\cite{HMF}
Throughout this paper, both the particle number $N$ and the simulation time $t$ are assumed to be $10^4$, and the initial distributions are the water-bag type.
We choose the phase constant of the one-body mean-field potential to be $\Phi(\theta,M)=1-M\cos \theta$, where $M$ denotes the magnetization of the system.
The one-particle energy function is $\varepsilon(\theta,p,M)=p^2/2+\Phi(\theta,M)$, and the self-consistency condition on the energy distribution $f(\varepsilon)$ is 
\begin{equation}
M=\frac{1}{N}\int d\theta dp \cos \theta f(\varepsilon(\theta,p,M))\;.\label{eq:sc}
\end{equation}

Now, we describe the formation process of the core-halo structure of an HMF system.
The formation process consists of four steps.
First, as the result of a trigger, which will create a chemical potential gap between the core and the halo, by the parametric resonance of the system with the initial strong oscillation of the magnetization, the core-halo structure starts forming\cite{PL}.
Second, after a {\it{dynamical}} process facilitated by particle and energy exchanges between the core and the halo, the distribution relaxes to a {\it{steady}} superposition of two components: that is, the core and the halo.
[Note that due to its long-range nature, the potential $\Phi(\theta,M)$ is common to the core and halo distributions.]
Here, we denote the fine-grained core and halo distributions by $f_c$ and $f_h$, respectively.
The dynamical relaxation between the core and the halo
\begin{equation}
df_a/dt\to 0\;,\ a=c,h\label{eq:dr}
\end{equation}
plays the role of the Vlasov fluid property of incompressibility for each component $a=c,h$.
This relaxation converges the total mass $N_a$ and the diluted phase-space density $\eta_a$ for each $f_a$ under the condition $\eta=\eta_c+\eta_h$, where $\eta$ denotes the fine-grained phase-space density of the system.
Thirdly, the magnetization stabilizes, and the system enters the QSS regime.
Finally, phase-mixing converges.
That is, $f$ and $f_a$ closely approximate functions of $\varepsilon$ only.
Then, due to Eq.(\ref{eq:dr}), $\pa f_a(\varepsilon) /\pa t\to 0$ holds.
Consequently, the total energy $E_a$ of each $f_a$ converges.
At this time, the core-halo formation is complete.

Based on this process, we derive the core-halo QSS distribution and its corresponding entropy, by following the discussion of collisionless ergodic relaxation by Lynden-Bell.\cite{LB}
The phase space is divided into macro-cells, that is, assemblies of micro-cells, and incompressible Vlasov elements occupy micro-cells.
From now on, while $\eta$ denotes the fine-grained phase-space density, we consistently denote the {\it{coarse-grained}} (macro-cell level) core and halo distributions by
\begin{eqnarray}
{f}_c(\theta_i,p_i)&=&{f}_{c,i}=\frac{\eta m_i \omega}{\nu\omega}=\frac{\eta_c m_i}{\nu_c}\;,\ \ \eta_c=\frac{\eta}{\nu}\nu_c\;,\label{eq:f1}\\
{f}_h(\theta_i,p_i)&=&{f}_{h,i}=\frac{\eta n_i \omega}{\nu\omega}=\frac{\eta_h n_i}{\nu_h}\;,\ \ \eta_h=\frac{\eta}{\nu}\nu_h\;,\label{eq:f2}
\end{eqnarray}
where $i$ labels macro-cells ($i=1,2,\ldots,P$), $m_i$ and $n_i$ are the numbers of Vlasov elements occupying the $i$-th macro-cell, $\nu$ is the number of micro-cells in each macro-cell, and $\omega$ is the area of each micro-cell.
In the process described above, the following partitions are fixed:
\begin{equation}
N=N_c+N_h\;,\ E=E_c+E_h\;,\ \nu=\nu_c+\nu_h\;.\label{eq:constraint}
\end{equation}
The third partition in Eq.(\ref{eq:constraint}) is kept for the ratios in the continuum limit $\nu\to 0$.
The total partition number of the configurations of Vlasov elements in the phase space is
\begin{eqnarray}
W&=&W_{mix} W_{lb}^{(c)} W^{(h)}_{lb}\;,\label{eq:W}\\
W_{mix}&=&\frac{N!}{N_c!N_h!}\prod_{i=1}^P\frac{\nu!}{\nu_c!\nu_h!}\;,\\
W_{lb}^{(c)}&=&\frac{N_c!}{\prod_{i=1}^P m_i!}\prod_{i=1}^P\frac{\nu_c!}{(\nu_c-m_i)!}\;,\\
W_{lb}^{(h)}&=&\frac{N_h!}{\prod_{i=1}^P n_i!}\prod_{i=1}^P\frac{\nu_h!}{(\nu_h-n_i)!}\;,
\end{eqnarray}
where $W_{mix}$ is the partition number of mixing the core and the halo and $W_{lb}^{(a)}$ are the Lynden-Bell partition numbers for the core and the halo.
Using Eqs.(\ref{eq:f1}) and (\ref{eq:f2}), the total partition number $W$ can be expressed as a functional of coarse-grained distributions ${f}_{c,i}$ and ${f}_{h,i}$.
The procedure for the maximization of the entropy in terms of these distributions is
\begin{equation}
\delta_{{f}_{c,i}} \ln W=0\;,\ \ \delta_{{f}_{h,i}}\ln W=0
\end{equation}
under the constraints in Eq.(\ref{eq:constraint}).
We introduce two kinds of Lagrange multiplier, $\alpha_a$ and $\beta_a$, where $a=c,h$, for fixed particle number $N_a$ and energy $E_a$, respectively.
Under the continuum limit ($\nu\to 0$), the total entropy reduces to
\begin{equation}
S=S^{(c)}+S^{(h)}\;,\label{eq:S}
\end{equation}
where each $S^{(a)}$ is the Lynden-Bell entropy\cite{LB}
\begin{equation}
S^{(a)}=-\int d\theta dp \biggl[\frac{f_a}{\eta_a}\ln \frac{f_a}{\eta_a}+\biggl(1-\frac{f_a}{\eta_a}\biggr)\ln\biggl(1-\frac{f_a}{\eta_a}\biggr)\biggr]\label{eq:LB}
\end{equation}
 for $a=c,h$.
The maximization solution of Eq.(\ref{eq:S}) is the double Lynden-Bell distribution
\begin{equation}
f(\theta,p)=\sum_{a=c,h}\frac{\eta_a}{\exp[\beta_a(\varepsilon(\theta,p,M)-\mu_a)]+1}\;,\label{eq:dfs}
\end{equation}
where $\mu_a=-\alpha_a/\beta_a$ is the chemical potential of the core or the halo.
With this, our scenario is complete.
Here, readers may think that since the partition number $W$ is the product in Eq.(\ref{eq:W}), the distribution function Eq.(\ref{eq:dfs}) would also be a product. 
However, the fine grains of the distribution functions $f_c$ and $f_h$ do not share any micro-cells.
Thus, $f$ is a superposition, that is, $f=f_c+f_h$.

\section{$N$-body simulation results}
In the $N$-body simulation described in the following sections, the initial phase-space distribution function $\hat{f}_0(\theta,p)$\footnote{In this paper, we use a hat to denote normalization by the factor $1/N$.} was the uniform water-bag type distribution over the rectangle $[-\theta_0,\theta_0]\times [-p_0,p_0]$, namely
\begin{equation}
\hat{f}_0(\theta,p)=\hat{\eta}\Theta(\theta_0-|\theta|)\Theta(p_0-|p|)\;,\label{eq:f0}
\end{equation}
where $\Theta$ is the Heaviside unit one-step function.
The parameters $\theta_0$ and $p_0$ of Eq.(\ref{eq:f0}) satisfy the relations $\sin\theta_0/\theta_0=M_0$, $p_0=\sqrt{6{\hat{E}}-3(1-M_0^2)}$, and $\hat{\eta}=1/(4\theta_0p_0)$ for initial magnetization $M_0$ and energy per particle ${\hat{E}}$.
Using these relations, when we fix $\hat{\eta}$, we can deduce $\hat{E}$ from $M_0$.
In order to take advantage of the Vlasov incompressibility, that is, the dynamical conservation of $\hat{\eta}$, we classify simulation data by the common value of $\hat{\eta}$.
In this paper, we consider $\hat{\eta}=0.15$.

\begin{figure}[ht]
\caption{The $\hat{\eta}=0.15$ configuration of four data $M_0=0.53,M_{{\rm{min}}},0.72,0.78$ on the $(M_0,{\hat{E}})$ plane to be used in Figs.3 and 4.
Here, $M_{{\rm{min}}}\sim 0.6556$ is the magnetization of the initial water-bag distribution Eq.(\ref{eq:f0}) at the minimum ${\hat{E}}$.
The blue curve and the red line represent the initial water-bag states $\hat{E}(M_0)$ and the energy per particle $\hat{E}_{\varepsilon_F}$ of the Vlasov stationary water-bag state $f_{\varepsilon_F}$ for ${\eta}=1500$, respectively.}
\includegraphics[width=0.66\hsize,bb=0 0 260 179]{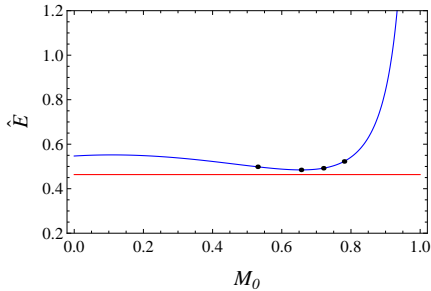}
\end{figure}

Figure 1 shows that $M_0=M_{{\rm{min}}}$ gives the global minimum of the function $\hat{E}(M_0)$ and in the neighborhood of this point, the function is convex.
This also holds for other values of $\hat{\eta}$.
Thus, it is natural to express some character of $f_0$ in terms of its total energy $E$.
Accordingly, we introduce the residual total energy ${{E}}_{{\rm{res}}}$ 
(refer to Fig.1)
which is equal to the total energy $E$ of the system minus the total energy ${E}_{\varepsilon_F}$ 
of the Vlasov stationary water-bag state $f_{\varepsilon_F}(\varepsilon)=\eta\Theta(\varepsilon_F-\varepsilon)$ for ${\eta}=1500$ (i.e., ${E}_{{\rm{res}}}={E}-{E}_{\varepsilon_F}$).\footnote{In this context, the Vlasov stationary water-bag state $f_{\varepsilon_F}$ depends on three parameters, that is, the Fermi energy $\varepsilon_F$, the magnetization and the total energy $E_{\varepsilon_F}$.
These are determined by the two conservation laws and the self-consistency condition Eq.(\ref{eq:sc}).}
The purpose of the introduction of $f_{\varepsilon_F}$ lies in its role in $E_{{\rm{res}}}$.
To show this, we note that $f_{\varepsilon_F}$ has a total energy lower than that of any $f_0$ with the common value of $\eta$ and cannot be accessible by Vlasov dynamics starting from $f_0$ due to energy conservation.
To clarify the meaning of $E_{{\rm{res}}}$, we consider the dynamics of the system on the phase space.
When the dynamics start from $f_0$, its center $f_{\varepsilon_F}$ is Vlasov stationary, and there is a total energy gap ${E}_{{\rm{res}}}>0$ between them.
So, by using $E_{{\rm{res}}}$ the system creates the halo of high-energy particles in the outer site, then, the inner part approaches the Vlasov stationary water-bag state due to energy conservation.
Thus, ${{E}}_{{\rm{res}}}$ measures the degree of the creation of the high-energy tail of the halo, which causes the system to deviate from the Lynden-Bell equilibrium.
That is, we argue that $E_{{\rm{res}}}$ is an {\it{a priori}} measure of the deviation of the system from the Lynden-Bell equilibrium.
For $M_0\le M_{{\rm{min}}}$, the Vlasov stationary water-bag distributions that inscribe and circumscribe the initial distribution Eq.(\ref{eq:f0}) are close to each other.
So, in these cases, the validity of this argument weakens.
In Fig.2, we illustrate this argument by the almost monotone correspondence between the residual energy per particle and the residue of the Lynden-Bell entropy of the Lynden-Bell equilibrium against that of the system.
\begin{figure}[ht]
\caption{This figure shows the residue of the Lynden-Bell entropy of the Lynden-Bell equilibrium $S_{{\rm{eq}}}$ for given ${{E}}$ and ${\eta}=1500$ against that of the simulation result $S_{{\rm{sim}}}$ as a function of the residual energy per particle ${\hat{E}}_{{\rm{res}}}$. 
Blue (upper) and purple (lower) plotted dots represent, respectively, the cases of $M_0=0.66\sim 0.78$ (with $0.01$ increments) and $M_0=0.41\sim 0.65$ (with $0.02$ increments) using the distributions averaged over $10$ runs.}
\includegraphics[width=0.66\hsize,bb=0 0 260 179]{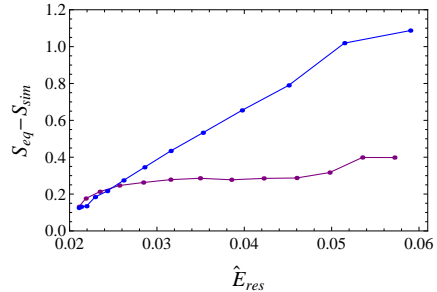}
\end{figure}
As easily confirmed, the minimization condition on the residual total energy $({\pa {{E}}}_{{\rm{res}}}/{\pa M_0})_{{\eta}}=0$ matches the GVC for the HMF model discussed in Ref.16.
So, our argument has an advantage over the GVC formulation.

As illustrated in Fig.3, as the residual total energy ${{E}}_{{\rm{res}}}$ increases, the high-energy tail of the simulation resultant $f(\varepsilon)$ grows.
This high-energy tail causes the simulation resultant $f(\varepsilon)$ to deviate from the Lynden-Bell equilibrium.
\begin{figure}[ht]
\caption{These figures show the deviation of the simulation resultant $f(\varepsilon)$ (red dots) averaged over $20$ runs from the single Lynden-Bell equilibrium (gray curve) for $M_0=M_{{\rm{min}}},0.72,0.78$ (from top to bottom).}
\includegraphics[width=0.5\hsize,bb=0 0 260 179]{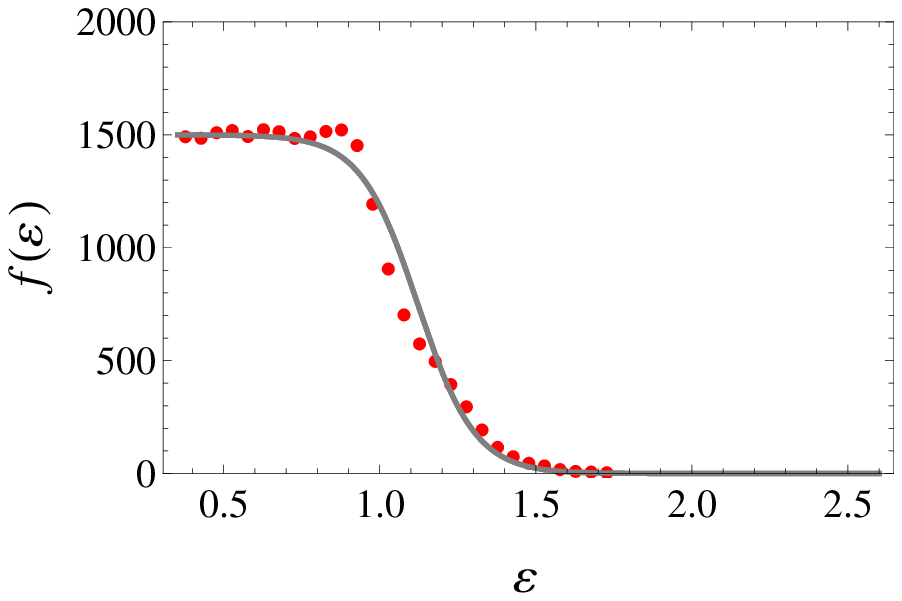}\includegraphics[width=0.5\hsize]{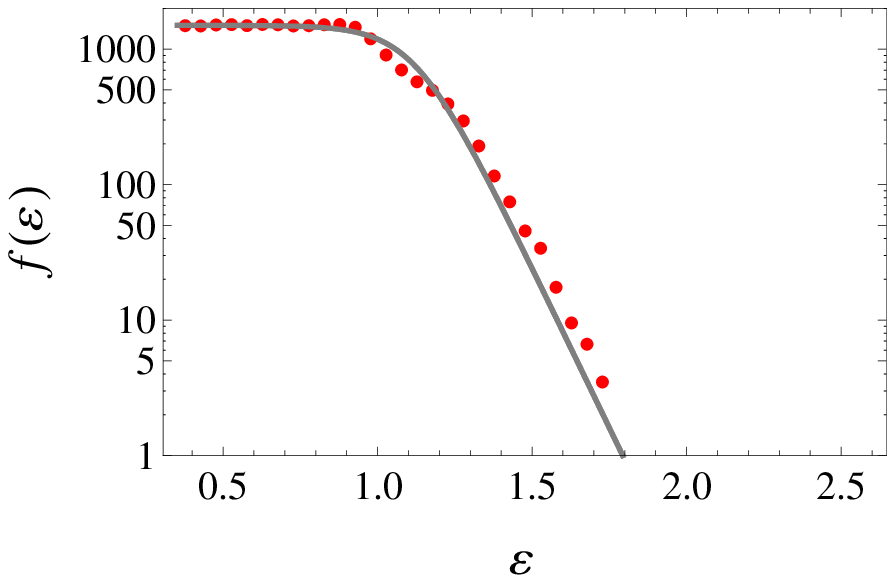}\\
\includegraphics[width=0.5\hsize]{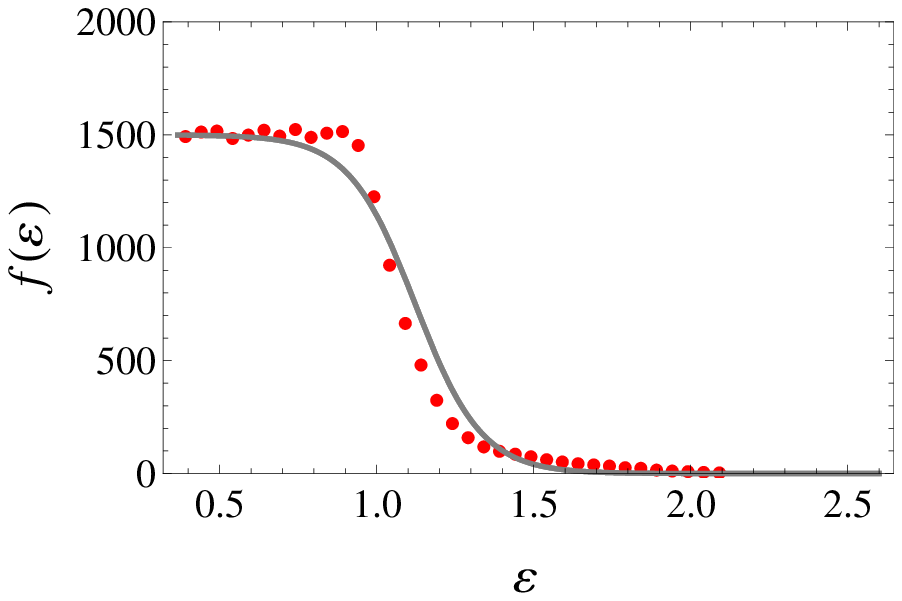}\includegraphics[width=0.5\hsize]{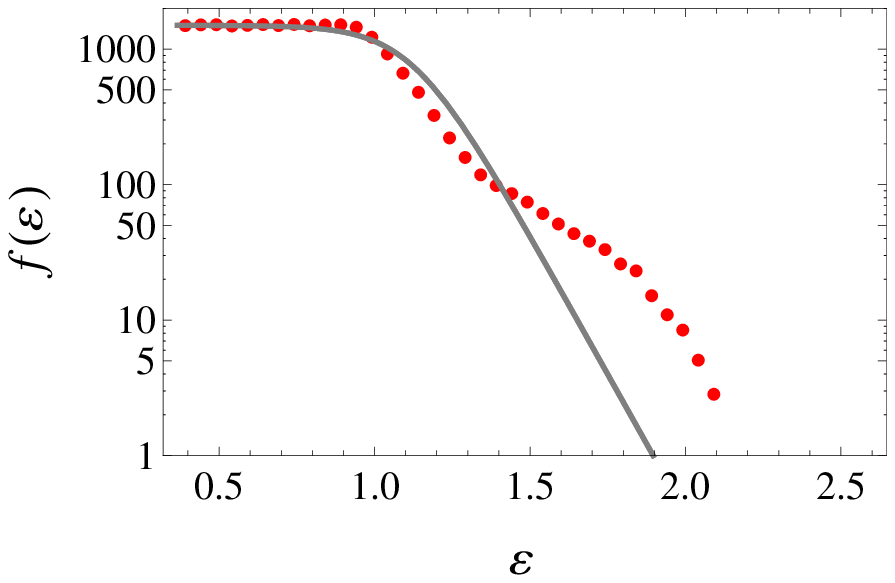}\\
\includegraphics[width=0.5\hsize]{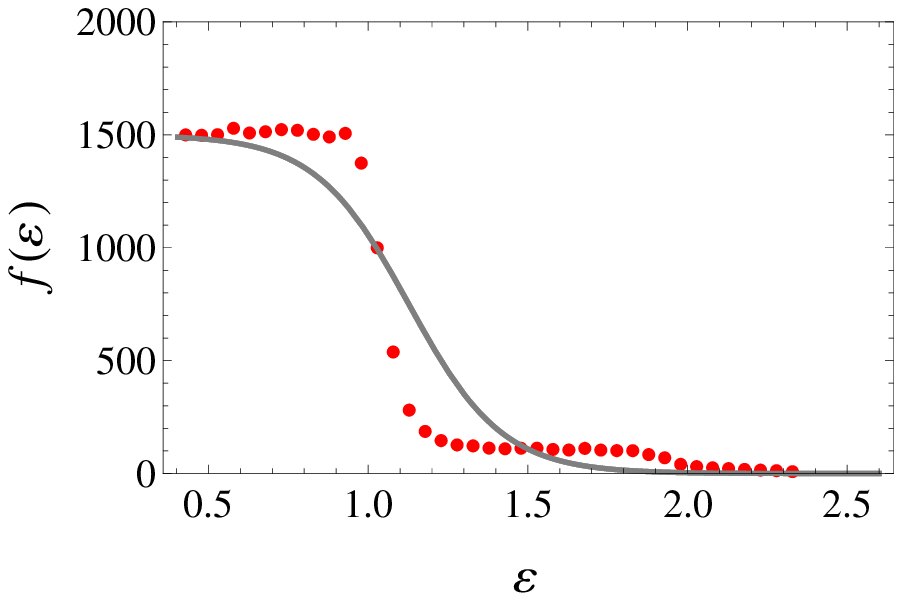}\includegraphics[width=0.5\hsize]{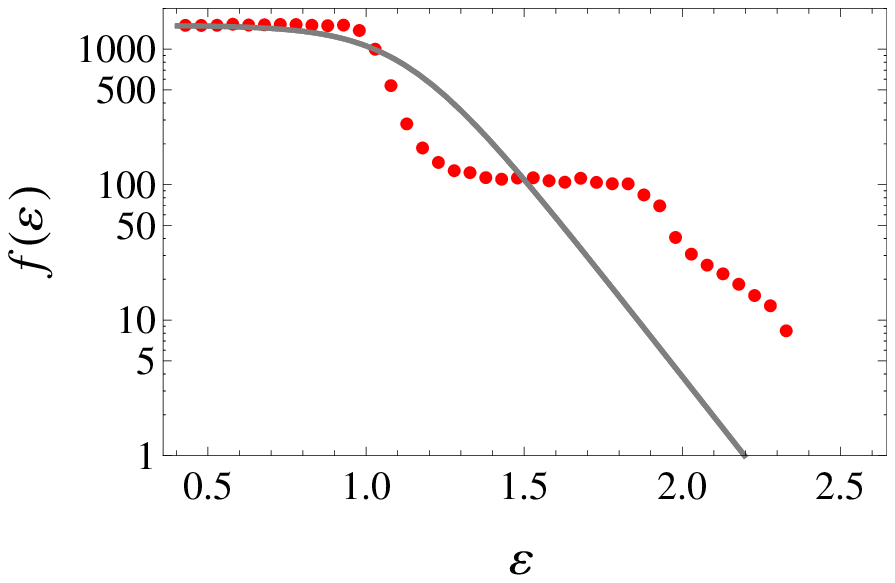}
\end{figure}
\begin{figure}[ht]
\caption{These figures show the $M_0=0.53,M_{{\rm{min}}},0.78$ (from top to bottom) double Lynden-Bell theoretical semi-predictions of simulation resultant $f(\varepsilon)$ (red dots) averaged over $20$ runs. Green (upper) and cyan (lower) solid curves represent the full and halo part of the double Lynden-Bell distribution, respectively. The former is constrained to satisfy the three conservation laws and the self-consistency condition Eq.(\ref{eq:sc}) and by adjusting values of the Lynden-Bell entropy, stationary magnetization $M_s$ and ${\eta}_c$ by hand. Gray dashed curves represent the Lynden-Bell equilibrium for given ${{E}}$ and ${\eta}=1500$.}
\includegraphics[width=0.5\hsize,bb=0 0 260 179]{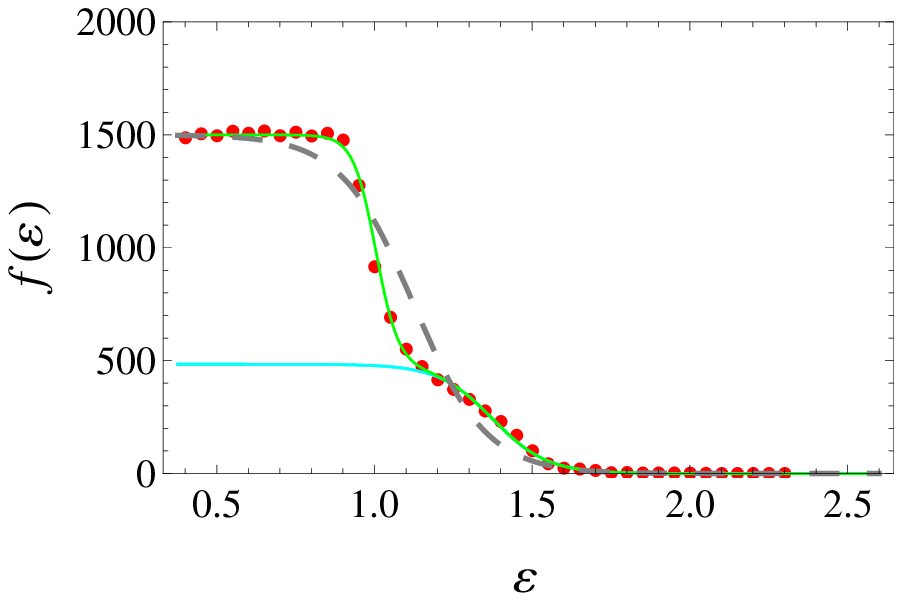}\includegraphics[width=0.5\hsize,bb=0 0 260 179]{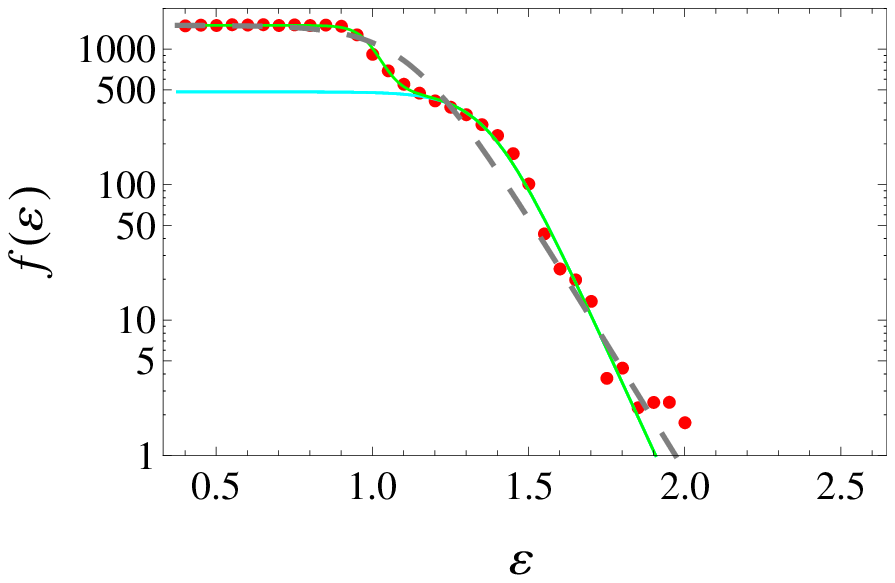}\\
\includegraphics[width=0.5\hsize,bb=0 0 260 179]{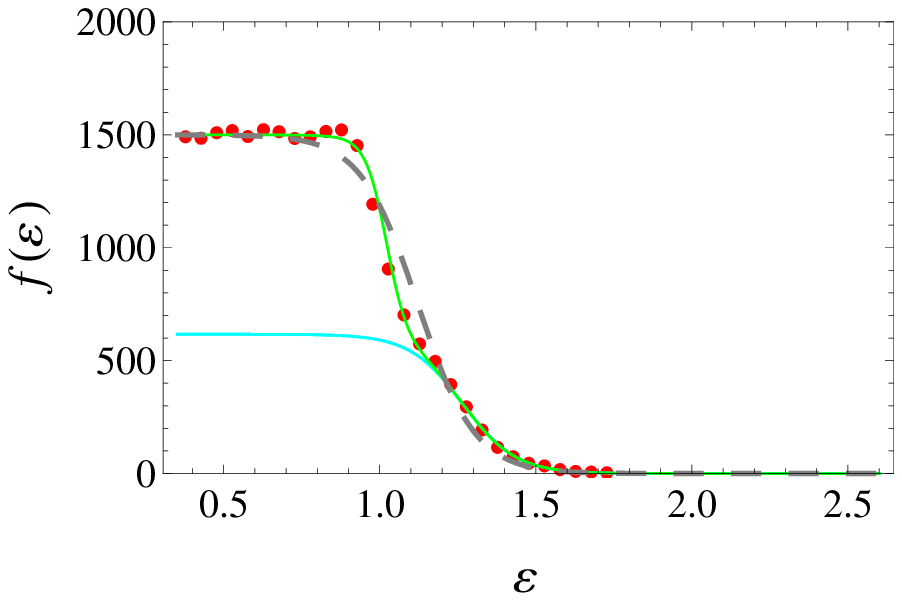}\includegraphics[width=0.5\hsize]{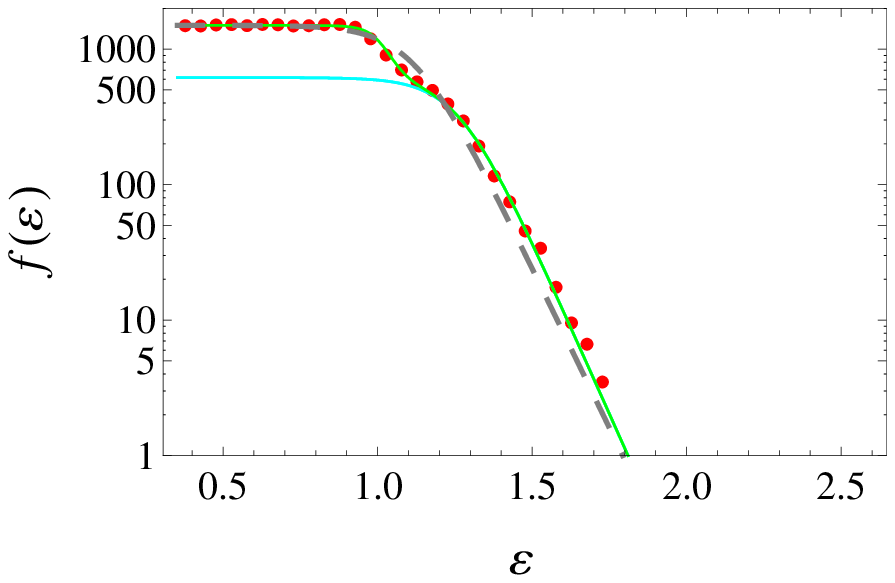}\\
\includegraphics[width=0.5\hsize]{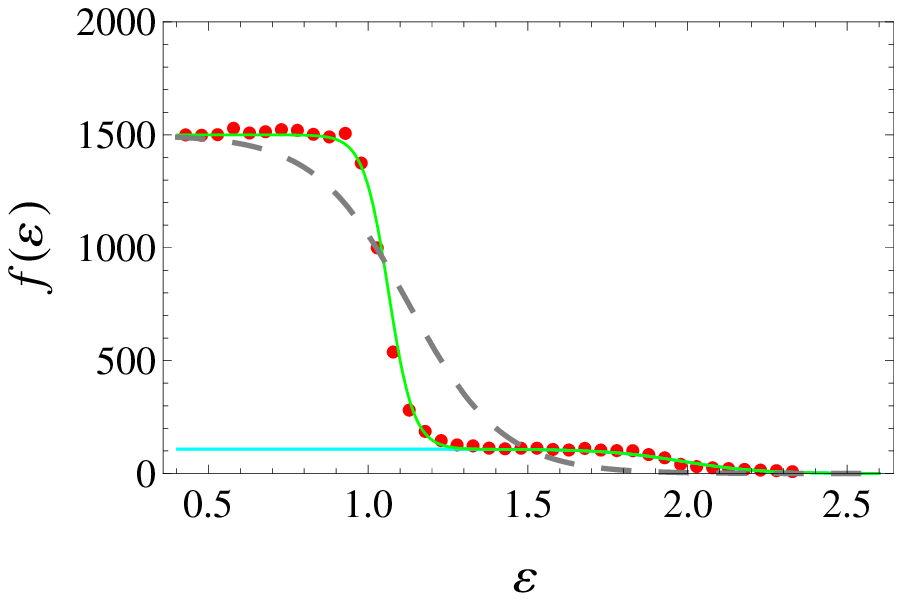}\includegraphics[width=0.5\hsize]{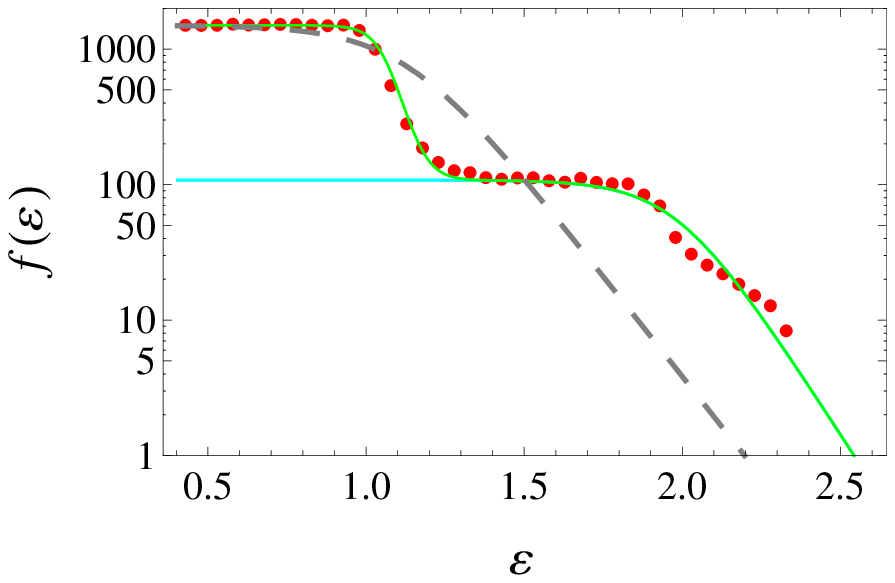}
\end{figure}
In the double Lynden-Bell scenario, we argue that this deviation part is fitted by the halo part of the distribution, $f_h$.
As an illustration of this argument, in Fig.4 we show the {\it{theoretical semi-predictions}} using the double Lynden-Bell distributions for the three initial magnetizations $M_0=0.53,M_{{\rm{min}}}$ and $0.78$.
[Note that a double Lynden-Bell distribution has seven degrees of freedom.
In Fig.4, by adjusting three parameters by hand, we solve the four conditions, that is, three conservation laws for mass, energy and phase-space density, and the self-consistency condition Eq.(\ref{eq:sc}), and derive the double Lynden-Bell distributions.
So, Fig.4 is not just a fitting but is also a theoretical result.]
The three parameters adjusted by hand to produce Fig.4 include the Lynden-Bell entropy.
By setting the Lynden-Bell entropy to be lower than that of the Lynden-Bell equilibrium, we accurately reproduce the $N$-body simulation results.
These accurate reproductions 
are pieces of corroborating evidence for
the double Lynden-Bell scenario.

\section{Relaxation criterion}
In this section, we examine whether or not the simulation resultant QSSs complete the relaxation between the core and halo Lynden-Bell distributions, which is a weaker criterion than Lynden-Bell relaxation.
The relaxation criterion to be considered can be expressed as the {\it{on-shell}} maximization of the double Lynden-Bell entropy Eq.(\ref{eq:S}):
\begin{equation}
\biggl[\frac{\pa S}{\pa \hat{X}_c}-\lambda \frac{\pa (M-M_s)}{\pa \hat{X}_c}\biggr]\biggl|_{M=M_s}=0\;,\ \ X=N,E,\eta\label{eq:OSEM1}
\end{equation}
for Lagrange multiplier $\lambda$, where the stationary magnetization $M_s$ is fixed by hand.
Equation (\ref{eq:OSEM1}) leads to
\begin{equation}
\frac{\pa S}{\pa \hat{N}_c}\biggl/\frac{\pa M}{\pa \hat{N}_c}=\frac{\pa S}{\pa \hat{E}_c}\biggl/\frac{\pa M}{\pa \hat{E}_c}=\frac{\pa S}{\pa \hat{\eta}_c}\biggl/\frac{\pa M}{\pa \hat{\eta}_c}\label{eq:OSEM}
\end{equation}
at $M=M_s$.

First, we explain Eq.(\ref{eq:OSEM}).
Due to the relation $S=S^{(c)}+S^{(h)}$, to calculate $\pa S/\pa \hat{X}_c$, it is sufficient to calculate $\pa S^{(a)}/\pa \hat{X}_a$:
\begin{eqnarray}
\frac{\pa S^{(a)}}{\pa \hat{N}_a}&=&\frac{\beta_a}{\hat{\eta}_a} \biggl(-\mu_a+\frac{\hat{V}_a}{\hat{N}_a}-\frac{\hat{N}_a}{2}\frac{\pa M_a}{\pa \hat{N}_a}M_s\biggr)\;,\label{eq:SN}\\
\frac{\pa S^{(a)}}{\pa \hat{E}_a}&=&\frac{\beta_a}{\hat{\eta}_a}\biggl(1-\frac{\hat{N}_a}{2}\frac{\pa M_a}{\pa \hat{E}_a}M_s\biggr)\;,\label{eq:SE}\\
\frac{\pa S^{(a)}}{\pa \hat{\eta}_a}&=&-\biggl(\frac{\pa S^{(a)}}{\pa \hat{N}_a}\frac{\hat{N}_a}{\hat{\eta}_a}+\frac{\pa S^{(a)}}{\pa \hat{E}_a}\frac{\hat{E}_a}{\hat{\eta}_a}\biggr)\;,
\end{eqnarray}
where we introduce
\begin{eqnarray}
V_a&=&\int d\theta dp\frac{\Phi(\theta,M_s)}{2}f_a(\varepsilon(\theta,p,M_s))\;,\\
M_a&=&\frac{1}{N_a}\int d\theta dp\cos\theta f_a(\varepsilon(\theta,p,M_s))\;. \label{eq:Mag}
\end{eqnarray}
Of course, $a=c,h$.
Note that these calculations are off-shell with respect to Eq.(\ref{eq:sc}).
By noting that $M=\hat{N}_cM_c+\hat{N}_hM_h$, calculations of the derivatives of the magnetization $M$ by the core's macro-variables $\hat{N}_c$, $\hat{E}_c$ and $\hat{\eta}_c$ are straightforward.
So, we omit these here.

Next, to discuss this criterion in the core's macro-variable space, we need to note that the off-shell existence region of the double Lynden-Bell distributions in $({\hat{N}}_c,{\hat{E}}_c,\hat{\eta}_c)$ space at given $M_s$, which we will call the {\it{double Lynden-Bell region}}, is not dense.
More precisely, on the $({\hat{N}}_c,{\hat{E}}_c)$ and $({\hat{N}}_c,\hat{\eta}_c)$ planes for fixed $\hat{\eta}_c$ and $\hat{E}_c$, respectively, the double Lynden-Bell region is restricted to a thin, spindle-shaped region and has the following two main structures.
First, it has two edges where the energy-distribution is the superposition of two Vlasov stationary water-bag distributions.
The boundaries connecting these edges represent states in which a part of the components has a Vlasov stationary water-bag distribution.
That is, at the edges and boundaries of the double Lynden-Bell region, the temperature of the corresponding component becomes zero (i.e., $\beta_a\to \infty$), so $\hat{f}_a(\varepsilon)$ reduces to $\hat{\eta}_a\Theta(\mu_a-\varepsilon)$.
Second, on the $({\hat{N}}_c,{\hat{E}}_c)$ plane, the center of this spindle-shaped double Lynden-Bell region is the off-shell maximization point of the double Lynden-Bell entropy Eq.(\ref{eq:S}) for fixed $\hat{\eta}_c$, namely $\pa S/\pa \hat{N}_c=\pa S/\pa \hat{E}_c=0$ holds.
 Using Eqs.(\ref{eq:SN}) and (\ref{eq:SE}), it can be shown numerically that the corresponding energy-distribution satisfies ${\beta_c}/{\hat{\eta}_c}\sim {\beta_h}/{\hat{\eta}_h}$ and $\mu_c\sim \mu_h$ and thus is almost a single Lynden-Bell one.

From Eq.(\ref{eq:OSEM}), we find that the on-shell entropy maximization criterion can be expressed geometrically as the tangency of contour surfaces of $S$ and $M$ in $({\hat{N}}_c,{\hat{E}}_c,\hat{\eta}_c)$ space.
The case $M_0=0.72$ fulfills this criterion for $M_s=0.6345$, which is within the oscillation range of the stationary magnetization (see Figs.5 and 6).

\begin{figure}[htb]
\caption{These figures show the $M_0=0.72$ theoretical result for $f(\varepsilon)$ obtained by solving the on-shell entropy maximization condition for $M_s=0.6345$, and the simulation resultant $f(\varepsilon)$ averaged over $20$ runs (red dots).
The full and halo part of the former are drawn as green (upper) and cyan (lower) curves, respectively.}
\includegraphics[width=0.5\hsize]{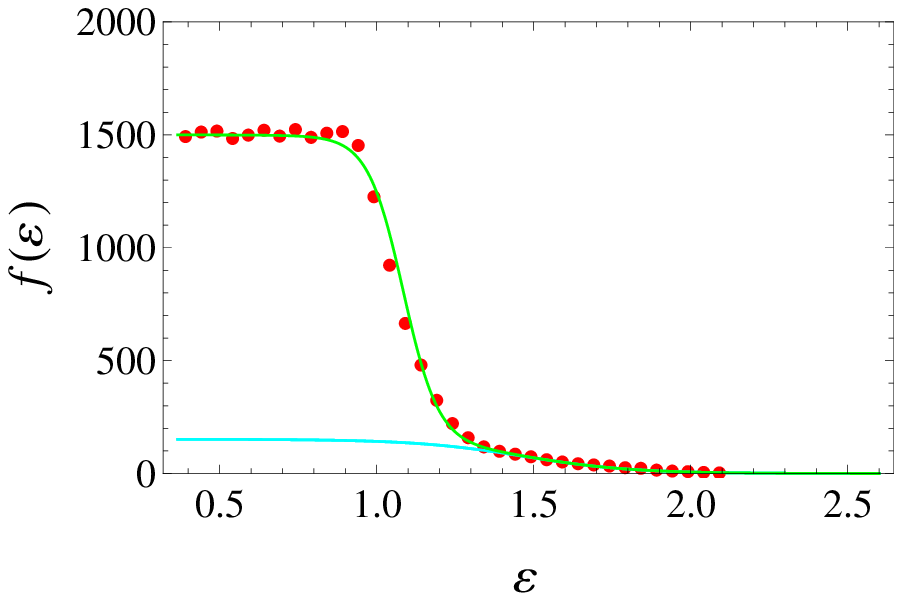}\includegraphics[width=0.5\hsize,bb=0 0 260 179]{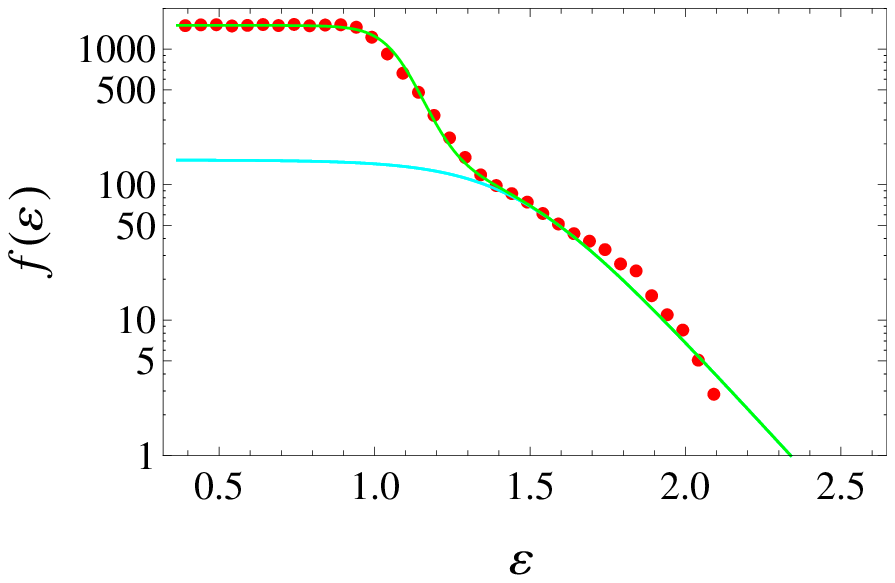}
\end{figure}

\begin{figure}[htb]
\begin{center}
\caption{These figures show $M_0=0.72$ contour curves of $S$ (purple, lower tangent) and $M$ (blue, upper tangent) on the $({\hat{N}}_c,{\hat{E}}_c)$ and $({\hat{N}}_c,\hat{\eta}_c)$ planes for $M_s=0.6345$ at the simulation resultants $\hat{\eta}_c$ and ${\hat{E}}_c$, respectively. The red dot is the simulation resultant point.
$S$ and $M$ are shown with equal contour intervals.}
\includegraphics[width=0.495325\hsize]{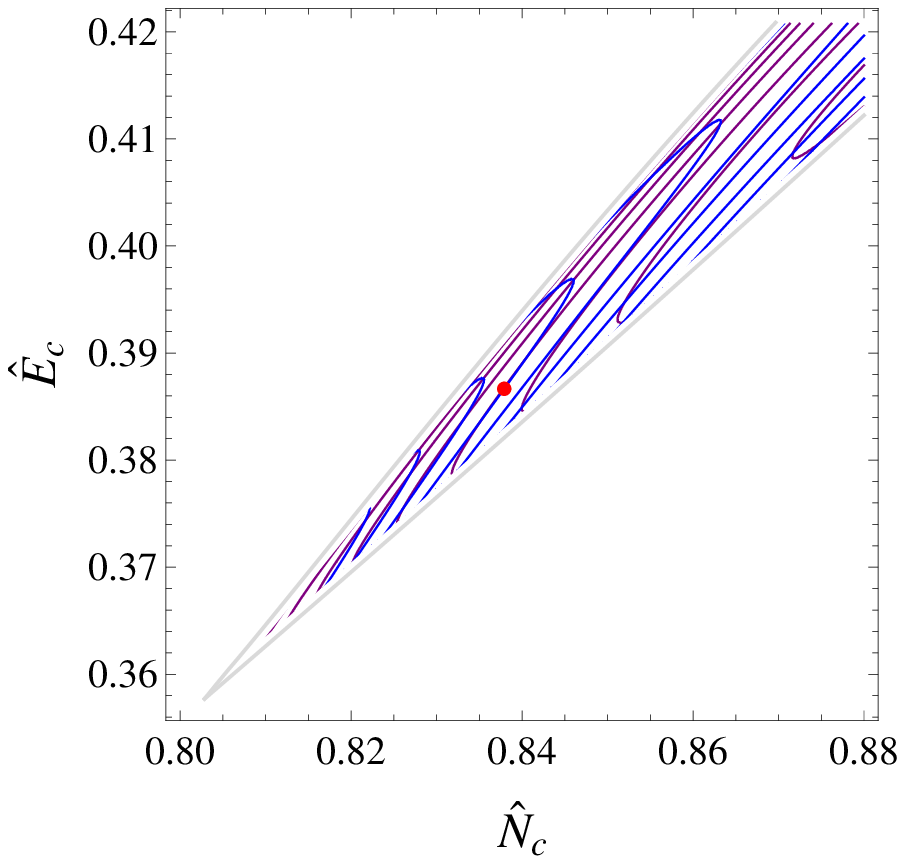}\includegraphics[width=0.504675\hsize]{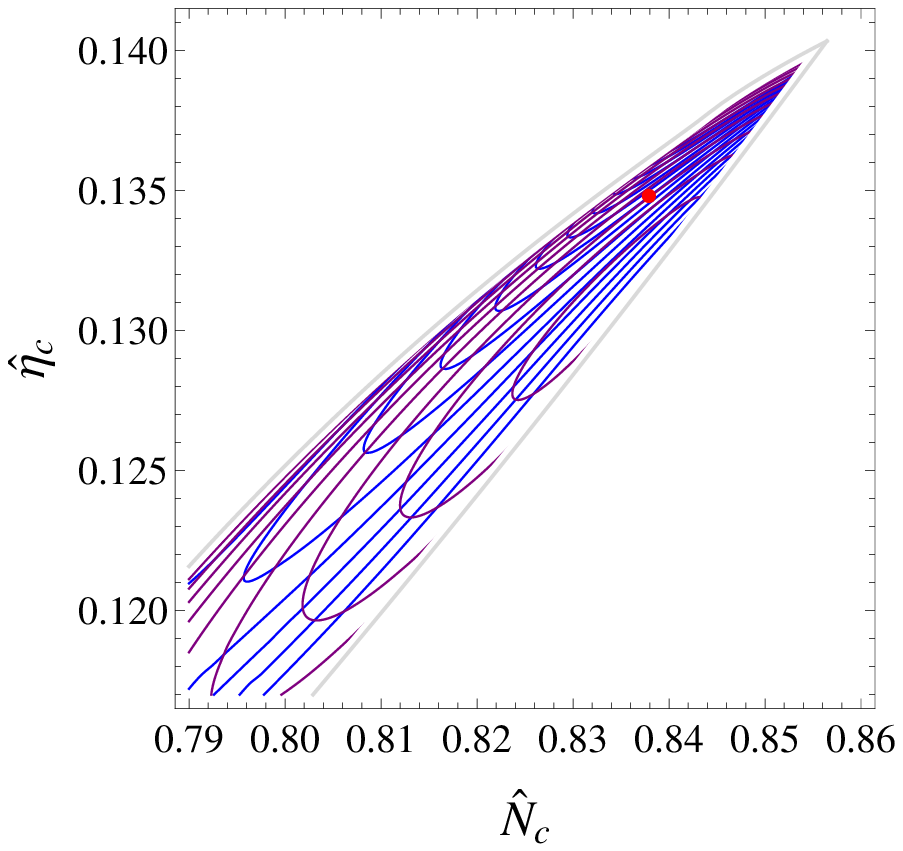}
\end{center}
\end{figure}

However, in other values of $M_0$, the simulation results do not fulfill the on-shell entropy maximization criterion and are regarded as cases of incomplete relaxation of Eq.(\ref{eq:S}).
This can be seen from the fact that the contours of $M$ on $({\hat{N}}_c,{\hat{E}}_c)$ plane, that is, slices of contour surfaces of $M$ at the simulation resultant $\hat{\eta}_c$, change from convex curves for $M_0=0.72$ to almost straight lines for other values of $M_0$, while the contour of $S$ is always convex.
In these cases, the tangent point between contours of $S$ and $M$ on this plane recedes, and the simulation results do not fulfill the on-shell entropy maximization criterion.

\section{Summary and outlook}
In the HMF model, we have systematically studied the core-halo structure of the QSSs for the unsteady (i.e., $M_0\neq 0$ or Vlasov unstable; $\hat{E}\le 7/12$\cite{HMF}) initial rectangle water-bag distributions with $N=10^4$ and $\hat{\eta}=0.15$ by means of $N$-body simulation and corroborated the double Lynden-Bell scenario at low energies per particle.
We also examined the completeness of collisionless relaxation by considering two entropies.
By using the Lynden-Bell entropy, we found that the systems being considered do not reach equilibrium and for higher total energy the degree of incompleteness increases.
By using the double Lynden-Bell entropy, in the case of $M_0=0.72$, the system completes the relaxation; however, for other values of $M_0$ this does not happen.

We now indicate some open issues.
First an {\it{a priori}} measure of the deviation from complete relaxation between the core and halo Lynden-Bell distributions needs to be found.
Second, it is important to apply the double Lynden-Bell scenario to unsteady systems at high energies per particle and determine the limits of the application of this scenario.
The effect of varying the total mass from $10^4$ should also be investigated.

\begin{acknowledgments}
We are grateful to T. Tashiro and T. Tatekawa for valuable discussions, especially to T. Tatekawa for providing the Fortran code for the HMF simulation.
\end{acknowledgments}


\begin{thebibliography}{99}
\bibitem{Book}A. Campa, T. Dauxois, D. Fanelli, and S. Ruffo, {\it{Physics of Long-range Interacting Systems}} (Oxford University Press, Oxford, 2014).
\bibitem{ReviewI}A. Campa, T. Dauxois, and S. Ruffo, 
Phys. Rep. {\bf{480}}, 57 (2009).
\bibitem{BMR}J. Barr${\acute{{\rm{e}}}}$, D. Mukamel, and S. Ruffo, Phys. Rev. Lett. {\bf{87}}, 030601 (2001).
\bibitem{MRS}D. Mukamel, S. Ruffo, and N. Schreiber, Phys. Rev. Lett. {\bf{95}}, 240604 (2005).
\bibitem{BH}W. Braun and K. Hepp, Commun. Math. Phys. {\bf{56}}, 101 (1977).
\bibitem{ReviewII}Y. Levin, R. Pakter, F. B. Rizzato, T. N. Teles, and F. P. da C. Benetti, Phys. Rep. {\bf{535}}, 1 (2014).
\bibitem{LPT}Y. Levin, R. Pakter, and T. N. Teles, Phys. Rev. Lett. {\bf{100}}, 040604 (2008).
\bibitem{TLPR}T. N. Teles, Y. Levin, R. Pakter, and F. B. Rizzato, J. Stat. Mech. (2010) P05007.
\bibitem{TLP}T. N. Teles, Y. Levin, and R. Pakter, Mon. Not. R. Astron. Soc. {\bf{417}}, L21 (2011).
 \bibitem{LB}D. Lynden-Bell, Mon. Not. R. Astron. Soc. {\bf{136}}, 101 (1967).
 \bibitem{HMF}M. Antoni and S. Ruffo, Phys. Rev. E {\bf{52}}, 2361 (1995).
\bibitem{PL}R. Pakter and Y. Levin, 
Phys. Rev. Lett. {\bf{106}}, 200603 (2011);
 L. D. Landau, J. Phys. USSR {\bf{10}}, 25 (1946).
 \bibitem{AFBCDR}A. Antoniazzi, D. Fanelli, J. Barr${\acute{{\rm{e}}}}$, P. H. Chavanis, T. Dauxois, and S. Ruffo, Phys. Rev. E {\bf{75}}, 011112 (2007).
 \bibitem{ACFR}A. Antoniazzi, F. Califano, D. Fanelli, and S. Ruffo, Phys. Rev. Lett. {\bf{98}}, 150602 (2007).
 \bibitem{AFRY}A. Antoniazzi, D. Fanelli, S. Ruffo, and Y. Y. Yamaguchi, Phys. Rev. Lett. {\bf{99}}, 040601 (2007).
\bibitem{BTPL}F. P. da C. Benetti, T. N. Teles, R. Pakter, and Y. Levin,
 Phys. Rev. Lett. {\bf{108}}, 140601 (2012);
  R. Bachelard, C. Chandre, D. Fanelli, X. Leoncini, and S. Ruffo, {\it{ibid}}. {\bf{101}}, 260603 (2008).
 \bibitem{Chavanis1}P. H. Chavanis, Physica. A {\bf{365}}, 102 (2006).
 \bibitem{CSR}P. H. Chavanis, J. Sommeria, and R. Robert, Astrophys. J. {\bf{471}}, 385 (1996).
\bibitem{Chavanis2}P. H. Chavanis, 
 in {\it{Multiscale Problems in Science and Technology}}, edited by N. Antonic {\it{et al.}} (Springer-Verlag, Berlin, 2002).
\end{thebibliography}
\end{document}